\documentclass[A4,12pt]{article}
\newcommand{\Ra}{\rangle}
\newcommand{\La}{\langle}

\newcommand{\T}{\mathbb T}

\newcommand{\Fr}{\frac{\hbar\omega}{2k_BT} }
\newcommand{\fr}{\frac{\hbar\omega}{2k_B} }
\newcommand{\Ch}{\mathop{\rm ch}}
\newcommand{\Sh}{\mathop{\rm sh}}
\newcommand{\Ct}{\mathop{\rm coth}  }
\newcommand{\Ash}{\mathfrak H}
\newcommand{\Mh}{\mathcal H}
\newcommand{\E}{\mathcal E}
\usepackage{amssymb,amsmath,   euscript}
\begin{document}
\begin{center}
{\large {\bf Quantum mechanical analogue \\of the zeroth law of thermodynamics~\footnote{The work was supported by RFBR (project no.12-01-90423)} }\\
 \small(On the problem of incorporating Thermodynamics into Quantum Theory})\\

\vspace{.23cm}
 {O.N. GOLUBJEVA$^\alpha$ and  A.D.SUKHANOV$^\beta$ }\\
\par\medskip
$^\alpha $ Peoples' Friendship University of Russia,  Moscow, Russia.\small{ ogol@mail.ru}\\

$^\beta$\small {BLTP, JINR, Dubna, Russia.}  \\
\end{center}
\begin{abstract}
This approach to the incorporation of stochastic thermodynamics into
quantum theory is based on the conception of consistent inclusion of
the holistic stochastic environmental influence described by wave
functions of the arbitrary vacuum, which was proposed in our paper
previously.

In this study, we implement the possibility of explicitly
incorporating the zeroth law of stochastic thermodynamics in the
form of the saturated Schr\"odinger uncertainty relation into
quantum theory.  This allows comparatively analyzing the sets of
states of arbitrary vacuums, namely, squeezed coherent states (SCSs)
and correlated coherent states (CCSs). In addition, we compare the
results of the construction of stochastic thermodynamics using SCSs
and TCCSs with the versions involving the incorporation of
thermodynamics into quantum theory developed previously and based on
thermofield dynamics and quantum statistical mechanics.
\end{abstract}


\section*{\small 1. Introduction }
The concept of thermal equilibrium in~\emph{thermodynamics} is
traditionally associated with the zeroth law, i.e., with the
equality (if only on the average) of Kelvin temperatures of the
object~$T$ and the environment~$T_0$, with which the model of
classical thermostat is brought in correspondence. Of course, the
concept of thermal equilibrium in the case of the~\emph{zero}
temperature is not introduced because the thermostat in the
traditional understanding (as a thermal environment) is not meant.
The tacitly taken thermal stochastic environmental influence is not
taken into account under these conditions, which are typical of
quantum mechanics.

At the same time, in the case of sufficiently low temperatures where
the thermal influence must be already  taken into account, the
quantum stochastic influence continues to be a significant factor.
Thus, quantum and thermal fluctuations are produced simultaneously
but they turn out to be nonadditive. We introduced [1] the concept
of quantum thermostat (arbitrary vacuum in the quantum language) as
a model of environment  to describe such a case at low temperatures.
To extend the concept of thermal equilibrium to the case of the
contact with the quantum thermostat, proceeding from empirical
considerations, we previously introduced the effective temperature
[2]
\begin{equation}\label{1}
\T=\fr\Ct{\Fr},
\end{equation}
where $k_B$ is the Boltzmann constant and $T$ is the Kelvin
temperature, and proposed the generalized zeroth law in the form
\begin{equation}\label{2}
\T=\T_0\pm\Delta\T,
\end{equation}
where $\mathbb T_0$ is the effective temperature of the quantum
thermostat and~$\Delta\T$ is the standard deviation of the effective
object temperature. In the limit when the Kelvin temperature tends
to zero~$T\rightarrow 0$, the minimum effective temperature for the
normal  vacuum mode (with the frequency~ $\omega$)
$\T_0^{min}\equiv\fr\ne0$. In this case the expression (2) takes the
form
\begin{equation}\label{3}
\T^{min}=\T_0^{min}\pm\Delta\T^{min},
\end{equation}
and retains the meaning of the condition of equilibrium with the
\emph{cold vacuum}.

We now consider whether expression~ (2) can be treated fundamentally
on the microlevel. To do this, we turn to the Hamiltonian of the
arbitrary vacuum
\begin{equation}\label{4}
\hat{\mathfrak H}_{\tau,\varphi}=\hat{\mathcal
H}_{\tau,\varphi}-\hat{\mathfrak H}^{inf}_{\tau,\varphi},
\end{equation}
which we introduced in~[3]. In this expression, $\hat{\mathcal
H}_{\tau,\varphi}$ is the Hamiltonian of the~\emph{system} that is
modeled by a quantum oscillator with the frequency~ $\omega_0$, and
$\hat{\mathfrak H}^{inf}_{\tau,\varphi}$ is the~\emph{energy
operator} of the holistic ( i.e. quantum and thermal) stochastic
environmental \emph{ influence} at an arbitrary Kelvin temperature.
We recall that the subscripts~$\tau$ and~$\varphi$ are the
parameters~$(u,v)$ of the Bogolyubov transformation allowing us to
pass from the~\emph{initial} (cold) vacuum to an~\emph{ arbitrary}
one (including the thermal one):
\begin{equation}\label{5}
u=\Ch\tau\cdot e^{i\varphi};\;\;\;\; v=\Sh\tau\cdot e^{-i\varphi}.
\end{equation}
The system Hamiltonian~$\hat{\mathcal H}_{\tau,\varphi}$ contained
in~(4) is expressed through these parameters in the following way:

\begin{equation}\label{6}
\hat{\mathcal{H}}_{\tau,\varphi} \equiv (\Ch 2\tau-\Sh
2\tau\cdot\cos 2\varphi)\hat K_{\tau,\varphi}+(\Ch 2\tau+\Sh
2\tau\cdot\cos 2\varphi)\hat\Pi_{\tau,\varphi},
\end{equation}
where~$\hat K_{\tau,\varphi}$ и $\hat\Pi_{\tau,\varphi}$ are the
respective kinetic and potential energy operators.

The second operator~$\hat{\mathfrak H}^{inf}_{\tau,\varphi}$ in~(4)
consists also of two terms reflecting the contributions of both the
purely quantum influence~$\hat j_0= \dfrac\hbar2\hat I$ of
the~\emph{cold vacuum} and the additional
influence~$\hat{\sigma}_{\tau,\varphi} $, which appears during the
passage from the~ \emph{initial} cold vacuum (i.e., at the
environmental temperature~$T=0$) to the~\emph{arbitrary} one
admitting cases with nonzero environmental temperatures. In this
case,
\begin{equation}\label{7-39}
\hat{\mathfrak{H}}^{inf}_{\tau,\varphi}=\omega \hat
j_0+\omega\hat{\sigma}_{\tau,\varphi}.
\end{equation}
Here,
\begin{equation}\label{8}
\hat{\sigma}_{\tau,\varphi}\equiv(\Sh2\tau\cdot\sin2\varphi)\frac
12(\hat p\,\hat q+\hat q\hat p).
\end{equation}
 For~$\tau=0$ and~$\varphi=0$, operator~(7), of course, transforms into the energy operator of the initial cold vacuum.

To calculate the average value of operator~(4), below we use the
complex wave function~$\psi_{\tau,\varphi}$ of the arbitrary vacuum.
In the coordinate representation, it has the form

\begin{equation}\label{9-29}
\psi_{\tau,\varphi}(q,\omega)=\left[2\pi(\ \Delta
q_{\tau,\varphi})^2\right]^{-1/4}\exp\left \{-\frac{q^2}{4(\Delta
q_{\tau,\varphi})^2}( 1-i \beta_ {\tau,\varphi})\right\},
\end{equation}
where  $(\Delta q_{\tau,\varphi})^2$ is the coordinate variance for
the arbitrary vacuum and $\beta_{\tau,\varphi}$ is a significant
parameter determining the phase of the wave function. Function~(9)
is the eigenfunction of Hamiltonian~$\hat{\mathfrak
H}_{\tau,\varphi}$~(4) of the arbitrary vacuum with the  zero
eigenvalue.  [3]

\begin{equation}\nonumber
\;\;\;\;\hat{\mathfrak H}_{\tau,\varphi} \psi_{\tau,\varphi}
(q,\omega) =0\cdot\psi_{\tau,\varphi} (q,\omega) =0.
\end{equation}
The quantities~$(\Delta q_{\tau,\varphi})^2$
and~$\beta_{\tau,\varphi}$ in (9) are expressed in terms of the
parameters~$\tau$ and~$\varphi$ as follows:
\begin{equation}\label{10-30}
\begin{array}{l}
(\Delta q_{\tau,\varphi})^2=(\Delta q_0)^2(\Ch 2\tau-\Sh
2\tau\cdot\cos2\varphi);\;\;\;\;\; (\Delta q_0)^2=\dfrac{\hbar}{2\omega_0};\\
\beta_{\tau,\varphi}=\Sh 2\tau\cdot\sin 2\varphi.
\end{array}
\end{equation}

The momentum variance for the
state~$\big|\psi_{\tau,\varphi}(q,\omega)\rangle$ has the form
\begin{equation}\label{11}
(\Delta p_{\tau,\varphi})^2=(\Delta p_0)^2(\Ch 2\tau+\Sh
2\tau\cdot\cos2\varphi),\;\;\;\;\mbox{где}\;\;\;\;(\Delta
p_0)^2=\dfrac{\hbar\omega_0}{2}.
\end{equation}

In this paper, we return to the study of criteria that allow
separating states corresponding to the thermal equilibrium with the
arbitrary vacuum from the set of states~$\psi_{\tau,\varphi}
(q,\omega)$ generated by the Bogolyubov $u,v$~transformation.
Squeezed coherent states (SCSs) and correlated coherent states
(CCSs) are candidates for this role because the expressions for the
internal energy of the quantum oscillator in these states can be
reduced to the form coinciding with the Planck formula $\mathbb
U_{\scriptscriptstyle
T}=\frac{\hbar\omega}{2}\coth\frac{\hbar\omega}{2k_BT},$ where $k_B$
is the Boltzmann constant and $T$ is the Kelvin temperature. To do
this, it is possible to bring the parameter~ $\tau$ formally in
correspondence with the temperature in accordance with the relations
\begin{equation}\label{12}
\Ch 2\tau\equiv \Ct(\frac{\hbar\omega }{2k_BT} );\;\;\;\;\;\Sh
2\tau\equiv (\Sh\frac{\hbar\omega }{2k_BT})^{-1}
\end{equation}
 and to convince ourselves that  the obtained results and the Planck ones are similar in appearance. However, the fundamental concept of temperature has no pre-image in the microworld and is related to the concept of thermal equilibrium. Thus, the procedure for introducing another temperature-dependent parameter instead of~$\tau$ does not yet allow drawing the final conclusion on the equilibrium character of the given states and on the possibility of interpreting them as analogues of the thermal ones. Thus, it is required to formulate, in the quantum language, the condition that is analogous to that of thermal equilibrium with the thermostat at the temperature~$T$ adopted in thermodynamics and to verify the fulfillment of this condition in the cases of the SCSs and CCSs.

\section*{\small 2. Analysis of squeezed coherent states of the arbitrary vacuum }
In the paper~[4], we preliminary indicated that SCSs can be regarded
as thermal ones only  conditionally, keeping in mind that in
thermodynamics, only cases of the contact with the cold vacuum
(i.e., at the Kelvin temperature~$T=0$) can be brought in
correspondence to them. Based on the study of the form of the
Schr\"odinger uncertainty relations for these states, we now adduce
additional arguments in favor of this statement. In the case of the
SCS where $\varphi=0$, wave function~(9) becomes real because
$\beta_{\tau,\varphi}$ vanishes, and we let
$\psi_{\tau,0}\equiv\psi_0$ denote it. Substituting relations~(12)
in formula~(6), we formally change the parametrization of the system
Hamiltonian. However, we show below that this procedure turns out to
be meaningless in spite of the fact that we introduce another
notation~$\hat{\mathcal H}_{\scriptscriptstyle T}
 \equiv\hat{\mathcal H}_{\tau,\scriptscriptstyle 0}  $ for it instead
of~$\hat{\mathcal H }_{\tau,\varphi}  $. Indeed, assuming that the
oscillator mass is $m=1$, we obtain
 \begin{equation}\label{13}
 \;\;\;\hat{\mathcal H}_{\scriptscriptstyle T}
=[\Ct\Fr-(\Sh\Fr)^{-1}]\frac12\hat
p^2+[\Ct\Fr+(\Sh\Fr)^{-1}]\frac12\omega^2\hat q^2
\end{equation}
in this case. Now the operator~$\hat{\sigma}_{\tau,
0}\equiv(\Sh2\tau\cdot\sin2\varphi)\frac 12(\hat p\,\hat q+\hat
q\hat p)\Big|_{\varphi=0} $ corresponding to the correlation between
coordinate and momentum fluctuations in formula~(7) vanishes because
$\sin2\varphi=0$. Thus, the stochastic influence is characterized
only by the operator~$\hat j_0$, which is typical of the initial
cold vacuum so that expression~(7) becomes simplified: $\mathfrak
H^{inf}_{\tau,\varphi}=\omega\hat j_0$.

If the fact that the average value of the vacuum
Hamiltonian~$\hat{\mathfrak H}_{\tau,\varphi}$ vanishes is taken
into account, the equality
\begin{equation} \label{14}
\;\;\;\;\;\La\psi_{\scriptscriptstyle T}\big|\hat{\mathcal
H_{\scriptscriptstyle T}}\big|\psi_{\scriptscriptstyle
T}\Ra=\La\psi_{\scriptscriptstyle T}\big|\omega\hat {j}_0
\big|\psi_{\scriptscriptstyle T}\Ra.
\end{equation}
can be obtained from formula~(4). We attract attention to the fact
that the left- and right-hand sides of~ (14) contain physically
different quantities: $\hat{\mathcal H_{\scriptscriptstyle T}}$
characterizes the system modeled by the quantum oscillator, and
$\hat j_0$, its environment. Thus, equality~(14) means that the
average energy of the system (oscillator) coincides with that of the
quantum environmental influence. This assertion formulated in the
quantum language is, in fact, equivalent to the zeroth law of
thermodynamics, but for the limiting case of the contact of the
system with the cold vacuum.

It follows from formulas~ (14) and~(13) that
\begin{equation} \label{15}
 \;\;\;\;\; \left[\coth^2( \Fr)-\left(\sinh^2( \Fr
)\right)^{-1}\right]\mathbb U_0=\frac{\hbar\omega}{2}.
\end{equation}

However, we note that the expression in brackets in~ (15) is
\emph{identically} equal to unity. In this case, $\mathbb
U_0=\La\psi_0\big|\hat{\mathcal
H_0}\big|\psi_0\Ra=\dfrac{\hbar\omega}{2}$ is the average energy of
the oscillator in the cold vacuum so that this expression is
independent of the parameter~ $T$. In view of this argument, the
parameter~$T$ cannot be treated as temperature. In other words, the
appropriateness of the change in the parametrization in form~(12)
turns out to be very doubtful, as was expected. At the same time,
this procedure allows obtaining certain useful information
from~(15). So, taking comments to formulas~(10) and~ (11) into
account, we can state that
$$\frac{\mathbb U_0}{\omega}=\Delta q_ 0\cdot\Delta p_0=   (\mathcal U\mathcal P)_0,$$
where the notation~$ (\mathcal U\mathcal P)_0$ is introduced for the
product of uncertainties~$\Delta p_0\cdot\Delta q_0$. Then
expression~(15) is, in fact, the partial realization of the
Schr\"odinger UR, i.e., the saturated Heisenberg UR
$$(\mathcal U\mathcal P)_0=\dfrac{\hbar}{2}\;\;\;\;\;\
;\;\;\;\;\;\;\;\;\;\;\;\;\;\;\;\;\;\;\;\;\;\;\;\;\;\;\;\;\;\;\;\;\;\;\;\;(*),$$
which is not related to thermal effects in any way.

Thus, the set of SCSs describes the specific case of the system
equilibrium only with the cold vacuum, i.e., at~$T=0$, which
corresponds to a "maximally"\; isolated (from the nonclassical point
of view) system, i.e., the system experiencing only stochastic
quantum influence. In other words, for the SCS at any \emph{nonzero}
Kelvin temperature having the standard physical meaning, it is
improper to put a question of the equilibrium in the sense that is
meant by the zeroth law of thermodynamics In this case, there is no
"warm" thermostat providing the thermal contact. This confirms the
conclusion that probably, it is no mere chance that Umezawa called
squeezed states not truly thermal ones but only thermal-like states.

Thus, we draw the conclusion in accordance to which the saturation
of the UR in the Heisenberg form~(*) in the quantum language is not
a sufficient condition for the description of the thermal
equilibrium at~$T\neq 0$. Below, we show that the saturated
Schr\"odinger UR plays this role.
\section*{\small 3.  Analysis of correlated coherent states of the arbitrary vacuum based on the Schr\"odinger UR }
For the CCSs, another case occurs because the parameter~$\varphi =
\pi/4$.  In this case, for the zero average value of vacuum
Hamiltonian ~(4), we obtain the relation
\begin{equation}\label{16}
\;\;\;\;\;\La\psi_{\tau,\varphi}\big |\Mh\big
|\psi_{\tau,\varphi}\Ra=\La\psi_{ \tau,\varphi}\big |\Ash^{inf}\big
|\psi_{\tau,\varphi }\Ra.
\end{equation}
In accordance with~(7) and~(8), the operator~ $\Ash^{inf}$ in it
contains two nonzero terms.  If the parametrization is changed now
and the temperature is introduced in accordance with formula~(12),
then expressions~(10) and~(11) for variances become
\begin{equation}\label{17}
(\Delta q_{\scriptscriptstyle T})^2=(\Delta q_0)^2\Ct(\Fr)
;\;\;\;\;\; (\Delta p_{\scriptscriptstyle T})^2=(\Delta
p_0)^2\Ct(\Fr).
\end{equation}

Accordingly, after such an operation, the operator of additional
influence~ $\hat{\sigma}_{\tau,\varphi} $ (8) corresponding to the
correlation between coordinate and momentum fluctuations transforms
(for~$\varphi=\frac\pi4$) to the form
\begin{equation}\label{18}
\hat\sigma_{\scriptscriptstyle
T}\equiv(\Sh2\tau\cdot\sin2\varphi)\frac 12(\hat p\,\hat q+\hat
q\hat p)=[\frac{\hbar}{2}\frac{1}{\Sh\Fr}]\frac 12(\hat p\,\hat
q+\hat q\hat p)
\end{equation}

 Then taking into account (6) and~(7) and also (17) and~(18), from~(16), we obtain
\begin{equation}\label{19}
\;\;\;\;\;\Ct(\Fr)[\frac{\hbar\omega}{2}\Ct(\Fr)]=\frac{1}{\Sh(\Fr)}[\frac{\hbar\omega}
{2}\frac{1}{\Sh\Fr}]+\frac{\hbar\omega}{2}.
\end{equation}
 If formulas~(18) are now taken into account and the notation~$(\Delta p_{\scriptscriptstyle T})^2(\Delta
q_{\scriptscriptstyle T})^2\equiv (\mathcal U\mathcal
P)_{\scriptscriptstyle T}^2$ for the product of variances is used
again, the~\emph{left-hand} side of expression~ (16) can be written
in the form of the following equality after the multiplication
by~$\hbar/2\omega$:
\begin{equation}\label{20}
\;\;\;\;[\frac{\hbar}{\omega}\Ct(\Fr)]^2=\left({\frac{\mathbb
U_{\scriptscriptstyle T}}{\omega}}\right)^2=(\mathcal U\mathcal
P)_{\scriptscriptstyle T}^2 ,
\end{equation}
where $\mathbb U_{\scriptscriptstyle T} $ is the Planck oscillator
energy.

Accordingly, the \emph{right-hand} side of~(16) after the
multiplication by~$\hbar/2\omega$  can be rewritten in the form that
is more convenient for the interpretation from the physical point of
view:

\begin{equation}\label{21}
\;\;\;\;\;[\frac{\hbar}{2}\frac{1}{\Sh\Fr}]^2+(\frac{\hbar}{2})^2=
(\frac\hbar2\cdot\coth\dfrac{\hbar\omega}{2k_BT})^2=
 \Big|\La\psi_{\scriptscriptstyle T}|\hat p\cdot\hat
q|\psi_{\scriptscriptstyle T}\Ra\Big|^2\equiv \mathbb
{J}_{\scriptscriptstyle T}^2
\end{equation}
where the quantity~ $\mathbb {J}_{\scriptscriptstyle
T}=\dfrac\hbar2\cdot\coth\dfrac{\hbar\omega}{2k_BT} $ has the
meaning of the holistic effective stochastic environmental
influence, to which both Hermitian
operators~$\hat\sigma_{\scriptscriptstyle T}$ and~$\hat j_0$
contribute.
 Thus, equality~ (17) typical of the arbitrary vacuum with the wave function~$\psi_{\scriptscriptstyle T}$ is nothing but the saturated Schr\"odinger uncertainty relation (SUR)
\begin{equation}\label{22}
\;\;\;\;\; (\mathcal U\mathcal P)^2_{\scriptscriptstyle
T}=\sigma^2_{\scriptscriptstyle T}+\frac{\hbar^2}{4}
\end{equation}

At the same time, the direct comparison of the right-hand sides of
formulas~(20) and~(21) allows representing formula~ (17) in the form
\begin{equation}\label{23}
(\mathcal U\mathcal P)_{\scriptscriptstyle T} =\mathbb
{J}_{\scriptscriptstyle T}.
\end{equation}

In this case, we stress that the left- and right-hand sides of
formula~ (23) determined by expressions~(20) and~(21) have different
physical meanings in spite of the coincidence between their
dimensionalities. The expression~$(\mathcal U\mathcal
P)_{\scriptscriptstyle T} $ can be interpreted as a system
response~$\mathbb J_{\scriptscriptstyle {syst} }$ to the stochastic
environmental influence~ $\mathbb J_{\scriptscriptstyle T}$, so that
the equality between the influence on the system and its response
\begin{equation}\label{24}
\;\;\;\;\;\;\mathbb J_{\scriptscriptstyle {syst} }=\mathbb
J_{\scriptscriptstyle T}
\end{equation}
holds. We call the expression (24) \emph{the zeroth law of
stochastic thermodynamics}. Applying the general formula~
$\mathbb{J} =\frac{k_B}{\omega}\cdot\mathbb{T}$ for the effective
stochastic influence to~(24) on the left (for the system) and on the
right (for the quantum thermostat), we obtain the relation
\begin{equation}\label{25}
\mathbb T_{\scriptscriptstyle {syst} }=\mathbb T_{\scriptscriptstyle
T},
\end{equation}
which is the zeroth law of stochastic thermodynamics written in more
traditional form for \emph{effective} temperatures of the
environment and the object. Thus, saturated Schr\"odinger UR (SUR)~
(22) for thermal CCSs closely related to the fundamental description
on the microlevel acquires the status of the \emph{ quantum analogue
of the zeroth law of stochastic thermodynamics}. Therefore,
equality~ (24) is the essential requirement of the theory.

Concerning the zeroth law in form~ (24), we want to point out the
fact emphasizing the significant role of the phase of the wave
function. Indeed, the right-hand side of it contains the
contribution of the quantum thermostat influence
\begin{equation}\label{26}
\;\;\;\;\; (\mathbb J_{\scriptscriptstyle
T})=\sqrt{(\frac{\hbar}{2}\frac{1}{\Sh\Fr})^2+(\frac{\hbar}{2})^2}.
\end{equation}
The first term in it is determined by the average value of the
operator~$\hat\sigma_{\scriptscriptstyle T}$, which is dependent on
the phase of the wave function. It is not difficult to see that,
except for the case~$T\rightarrow 0$, the right-hand side of~ (26)
is mainly determined by precisely the phase of the wave function.
Its role is especially significant at high temperatures when the
corresponding term becomes
$$
(\frac{\hbar}{2}\frac{1}{\Sh\Fr})^2\rightarrow
(\frac{k_BT}{\omega})^2\gg(\frac{\hbar}{2})^2
$$

In this case, zeroth law~(24) itself reduces to the equality
condition for \emph{Kelvin}  temperatures~$T=T_0$, which is taken as
a definition of the thermal equilibrium in the classical
thermodynamics and in quantum statistical mechanics (QSM) (in this
case, the symbol~$T_0$ denotes the thermostat temperature). However,
it is important to stress that even in the cases where temperature
fluctuations of the system (consisting of many particles, for
example) can be neglected in the zeroth approximation, its
definition, which is adequate to this concept at the asymptotic, has
the form
\begin{equation}\label{27}
T\rightarrow\frac{\omega}{k_B}(\mathcal{U}\mathcal{P})_{\scriptscriptstyle
T}
\end{equation}
Thus, this quantity depends on the momentum and coordinate
uncertainties, which indicate its initially \emph{random} character.

In the cases where there is no phase, for example, when real wave
functions of the SCS type are used, formula~ (24) becomes
\begin{equation}\label{28}
\;\;\;\;\; \mathbb J_{syst}=\mathbb J_{qu},
\end{equation}
where $\mathbb J_{qu}=\dfrac\hbar2$ is the quantum influence. In
this case, the zeroth law is independent of the parameter~$T$, which
allows fixing the "thermal" equilibrium only with the cold vacuum,
as was mentioned above, i.e., treating it in the "Pickwick" sense.

Thus, to describe the thermal equilibrium with the arbitrary vacuum,
it is insufficient to have only two requirements, namely, the
definition of vacuum   and the coincidence between the expression
for the internal energy of the quantum oscillator formally expressed
in terms of the parameter~$T$ and the Planck formula. Both sets of
states (SCSs and CCSs) satisfy these requirements. However, only the
saturated Schr\"odinger UR corresponds to the inclusion of the
thermal influence on the microlevel. Therefore, it is possible to
endow the quantity~$T$ with the physical meaning that is adequate to
the concept of temperature only in the case of CCSs of the $\big
|\psi_{ \tau,\varphi}\Ra
  = \big |\psi_{ \tau,\frac\pi4}\Ra\equiv\big |\psi_{\scriptscriptstyle T}\Ra$ type,
which can be regarded as "\emph{thermal}" \;ones. In other words, to
separate the states characterizing the thermal equilibrium with the
arbitrary vacuum (for~$T\neq 0$) from all states generated by the
Bogolyubov $(u,v)$~transformations, one necessary requirement must
be introduced. It consists in that the equivalent of the zeroth law
of thermodynamics in the form of the saturated Schr\"odinger UR
\begin{equation}\nonumber
\Delta p \cdot\Delta q =\Big|\La\psi |\hat p\cdot\hat q|\psi
\Ra\Big|
\end{equation}
is included in the apparatus of quantum theory. In the state of the
"thermal" \;vacuum $\big|\psi_{\scriptscriptstyle T}(q)\Ra$, it
reduces to the condition
 \begin{equation}\nonumber
(\mathcal U\mathcal P)_{\scriptscriptstyle T}
=\frac\hbar2\coth\left( \frac{\hbar\omega}{2k_BT_0}\right),
\end{equation}
where we use~ $T_0$, which is the Kelvin thermostat temperature. We
stress that in this relation, the left-hand side is expressed in
terms of the object characteristics, and the right-hand side, in
terms of the environmental characteristics.

\section*{\small 4. Invariance of the zeroth law of stochastic thermodynamics}

We now analyze the above obtained fundamental formulation of the
zeroth law from the point of view of its invariance under the
Bogolyubov $(u,v)$~transformations. We attract attention again to
the fact that qualitatively different quantities are contained in
the right- and left-hand sides of the corresponding equalities;
however, their dimensionalities coincide.

In particular, formula~(28) means that at the zero temperature,
\begin{equation} \label{29}
\;\;\;\;\mathbb J_{qu}=\frac\hbar2= \Delta p_0\cdot\Delta
q_0\equiv(\mathcal U\mathcal P)_0,
\end{equation}
where $(\mathcal U\mathcal P)_0$ is regarded as a \emph{holistic}
quantity characterizing the object under study. We recall that
$(u,v)$~transformations retaining canonical permutation relations
have the meaning of transformations for two symmetry groups --- the
group~ $SU(1,1)$ and its locally isomorphic Lorentz group~ $O(2,1)$
--- simultaneously. Their common invariant is $\hbar^2/4$. We now
discuss how the presence of the corresponding invariance affects the
quantities contained in the left- and right-hand sides of
formula~(29). As follows from the preceding, under transformations
with the real parameters~ $u$ and~$v$ (i.e., for~$\varphi=0$ in
formula~(5)), the expression in the right-hand side of~(29) remains
the same because, in this case, the average value of the operator of
quantum thermal stochastic influence~$\hat\sigma$ remains equal to
zero for any values of the parameter~ $\tau$. At the same time, the
expression in the left-hand side of~ (29) remains invariant;
however, it can have different forms, depending on the choice of the
parameter~$\tau\neq 0$. In other words, the expression~$(\mathcal
U\mathcal P)^2$ in the general case can have the form
\begin{equation}\label{30}
\;\;\;\;\;(\mathcal U\mathcal P)^2=(\mathcal U\mathcal
P)^2_0\;(\cosh^{2} 2\tau- \sinh^{2} 2\tau)=(\frac\hbar2)^2,
\end{equation}
which corresponds to different ways of realizing SCSs.

If formulas~ (12) and also~(10) and~(11) are taken into account,
expression~(30) can be written as a determinant of the form
\begin{eqnarray}\label{31}
\;\;(\mathcal U\mathcal P)^2=(\mathcal U\mathcal P)^2_0
\begin{vmatrix} (\coth\frac{\hbar\omega}{2kT}+\frac{1}{\sinh\frac{\hbar\omega}{2kT}})&0\\
0&(\coth\frac{\hbar\omega}{2kT}-\frac{1}{\sinh\frac{\hbar\omega}{2kT}})
\end{vmatrix}=\nonumber\\=
\begin{vmatrix}
(\Delta p_{\scriptscriptstyle T})^2&0\\
0&(\Delta q_{\scriptscriptstyle T})^2
\end{vmatrix} =(\frac\hbar2)^2.
\end{eqnarray}
If, in this case, the momentum and coordinate variances are
calculated by averaging with the real wave function that follows
from formula~(9) for~$\beta_{\tau,\varphi}=0$, which is said to be
thermal-like, then (in accordance with the Umezawa terminology [5])
we deal with \emph{single-mode} squeezed states. Trying to
generalize formula~ (31), Umezawa  introduced two independent sets
of quanta, namely, \emph{ordinary} quanta related to the standard
operators of annihilation~$\hat a$ and creation~$\hat a ^+$ and
so-called \emph{thermal} quanta related to the operators~$\hat
{\tilde a}$ and~$\hat {\tilde a}^+$ with tildes. Because of this, he
doubled the Hilbert space, which allows him to introduce
\emph{two-mode} squeezed states. They are obtained using the
creation operators formed by means of the Bogolyubov
$(u,v)$~transformations entangling operators from two sets, which he
introduced, one with another. As a result, expression~ (30) becomes
\begin{equation}\label{32}
\;\;\;\;\;(\mathcal U\mathcal P)^2 =
\begin{vmatrix}
(\Delta\hat p)^2_{\scriptscriptstyle T}&(\Delta p\;\Delta\tilde p)_{\scriptscriptstyle T}\\
(\Delta q\;\Delta\tilde q)_{\scriptscriptstyle T}& (\Delta\hat
q)^2_{\scriptscriptstyle T}
\end{vmatrix}
=(\frac\hbar2)^2
\end{equation}
where the averaging is carried out over the thermal-like states, as
in~(31).

We stress that the off-main-diagonal terms in determinant~ (31) have
the forms of correlators of homogeneous quantities ($\hat
p,\hat{\tilde p}$) or ($\hat q,\hat{\tilde q}$); however, in this
case, correlators of heterogeneous quantities of the ($\hat p,\hat
q$) type, and so on, remain equal to zero. Thus, the right-hand-side
of equality~ (29) remains the same, which corresponds to the
property of the introduced states to be squeezed.

We note that the quantities contained in formulas~(31) and~(32) are
averaged using real wave functions. At the same time, in the
Umezawa's opinion, analogous results can also be obtained using the
real density matrix. Recently, Park made a successful effort of this
sort for the quantum oscillator in the thermal equilibrium [6].

He proposed a method for calculating the elements of the determinant
contained in~(32) using the averaging with the Gibbs--von Neumann
density operator
\begin{equation}\label{33}
\;\;\;\;\;\hat\rho_{\scriptscriptstyle T}=Z^{-1}\exp[-\Fr(\hat
N_a+\frac12\hat I)]
\end{equation}
where $Z=\mathrm{Sp}\exp[-\Fr(\hat N_a+\frac12\hat I)]$, $\hat I$ is
the unity operator.

The result obtained by Park allows writing determinant~(32=83) in
the form

\begin{equation}\label{34}
 \;\;\;\; \;\;\;\;\;(\mathcal U\mathcal P)^2=
\begin{vmatrix}
\mathrm{Sp}[(\Delta\hat p)^2\hat\rho_{\scriptscriptstyle T}] &
\mathrm{Sp}[\Delta\hat p\Delta\hat\rho_{\scriptscriptstyle
T}^{\frac12}
\Delta\hat p\Delta\hat\rho_{\scriptscriptstyle T}^{\frac12}] \\
\mathrm{Sp}[\Delta\hat q\Delta\hat\rho_{\scriptscriptstyle
T}^{\frac12} \Delta\hat q\Delta\hat\rho_{\scriptscriptstyle
T}^{\frac12}] & \mathrm{Sp}[(\Delta\hat
q)^2\hat\rho_{\scriptscriptstyle T}]
\end{vmatrix}
\end{equation}

We attract attention to the fact that, to provide the invariance
of~$\mathcal U\mathcal P$, the off-main-diagonal terms that are
absent in traditional approaches to the calculation of~ $\mathcal
U\mathcal P$ are added by Park. From the mathematical point of view,
the necessity of such a generalization is due to the fact that, in
this case, the operators~$(\Delta {\hat p})^2$ and~$\hat
{\rho_{\scriptscriptstyle T}}$ or $(\Delta {\hat q})^2$ and~$\hat
{\rho_{\scriptscriptstyle T}}$ under the trace sign do not commute
one with the other. Therefore, the method for calculating variances
adopted in the classical probability theory or in quantum mechanics
in the case where pure states are used requires generalization,
which was emphasized in the paper of Wigner and Yanase [7].

In accordance with Park, the off-diagonal terms in determinant~(34)
can be transformed using formula~(33) and the commutation rules
\begin{equation}\label{35}
\;\;\;\;\hat\rho_{\scriptscriptstyle{T}}^{\frac12}\;\hat
a=\exp(\Fr)\;\hat a\hat\rho_{\scriptscriptstyle
T}^{\frac12};\;\;\;\;\hat\rho_{\scriptscriptstyle{T}}^{\frac12}\;\hat
a^+=\exp(\Fr)\;\hat a^+\hat\rho_{\scriptscriptstyle T}^{\frac12}
\end{equation}

To do this, it is necessary to take into account that, in this case,
$\overline{(\Delta {\hat p})^2}=\overline{({\hat p})^2}$ and
$\overline{(\Delta {\hat q})^2}=\overline{({\hat q})^2}$ and to
express $\hat p$ and~$\hat q$ in terms of~$\hat a$ and~$\hat a^+$.

Thus, we obtain
\begin{equation}\label{36}
 \;\;\;\;\mathrm{Sp}[\hat
p\hat\rho_{\scriptscriptstyle{T}}^{\frac12}\hat
p\hat\rho_{\scriptscriptstyle{T}}^{\frac12}]=\mathrm{Sp}[\hat p\hat
p_{\scriptscriptstyle{T}}\hat\rho_{\scriptscriptstyle{T}}],
\end{equation}
where
\begin{equation}\label{37}
 \;\;\;\; \hat p_{\scriptscriptstyle
T}=\frac{i}{\sqrt{2}}(e^{\frac12\Fr}\hat a-e^{-\frac12\Fr}\hat a^+)
\end{equation}
and analogously
\begin{equation}\label{38}
\;\;\;\;\mathrm{Sp}[\hat
q\hat\rho_{\scriptscriptstyle{T}}^{\frac12}\hat
q\hat\rho_{\scriptscriptstyle{T}}^{\frac12}]=\mathrm{Sp}[\hat q\hat
q_{\scriptscriptstyle{T} }\hat\rho_{\scriptscriptstyle{T}}],
\end{equation}
where
\begin{equation}\label{39}
\;\;\;\; \hat q_{\scriptscriptstyle
T}=\frac{1}{\sqrt{2}}(e^{\frac12\Fr}\hat a+e^{-\frac12\Fr}\hat a^+)
\end{equation}

Expressions~(37 ) and~(39 ) can be regarded as renormalized momentum
and coordinate operators taking the thermostat influence into
account. The relation~$(\mathcal U\mathcal P)^2$ in the left-hand
side of~(34) expressed in terms of them can be written as the
determinant
\begin{equation}\label{40}
\;\;\;\;\;(\mathcal U\mathcal P)^2=
\begin{vmatrix}
(\Delta\hat p)^2_\rho&(\Delta p\;\Delta\tilde p)_\rho\\
(\Delta p\;\Delta\tilde p)_\rho& (\Delta\hat q)^2_\rho
\end{vmatrix}
\end{equation}
in which the calculation of all average quantities with the density
matrix~$\hat \rho_{\scriptscriptstyle T}$ leads to the left-hand
side of~(34). The comparison of~ (40) with the left-hand side of
expression~ (32) shows that the result of calculating~$(\mathcal
U\mathcal P)^2$ by Park using the real density matrix~$\hat\rho_T$
does not differs significantly from the similar result obtained by
Umedzawa using real two-mode squeezed states. However, it is
important to stress that both results~(32) and~(41) are related to
the case of the equilibrium of the system with the cold vacuum
because the right-hand side of zeroth law~(29) remains the same
during the averaging over real pure and mixed states. Thus, the
formal use of real states or real elements of the density matrix
does not allow describing the equilibrium of the system with the
thermal vacuum not only in the case of squeezed states, but also in
the cases of the Umedzawa thermofield dynamics and the QSM-based
thermodynamics even after the generalization of the latter in the
Wigner and Yanase spirit.

We now discuss the invariance of the zeroth law with respect to the
transitions between equilibrium thermal CCSs at nonzero
temperatures. To do this, we use the invariance of the zeroth law
with respect to the Lorentz group~$O(2,1)$. It can be demonstrated
in the simplest way using the right-hand side of condition~(29) by
replacing~$(\mathbb J_0^{inf})^2$ in accordance with formula~(21).
Then we obtain
\begin{equation}\label{41}
\;\;\;\;\;(\mathbb J_{qu})^2=(\mathbb J_{\scriptscriptstyle T}
)^2-\sigma^2_{\scriptscriptstyle T}=(\frac\hbar2)^2
\end{equation}

Hence it follows that for thermal CCSs, the set of
quantities~$\mathbb J_{\scriptscriptstyle T},\sigma
_{\scriptscriptstyle T}$ can be regarded as a two-dimensional
time-like vector in the pseudoEucledean space of states, and the
quantity~$\hbar^2/4$, as a length of this vector or the invariant of
the Lorentz group~$O(1,1)$ that is analogous to the eigentime in the
standard special theory of relativity in the world of events (the
group dimensionality decreases because the average value of the
Lagrangian is zero for thermal CCSs). The roles of the traditional
Lorentz factors~$\beta_{\scriptscriptstyle L}$
and~$\gamma_{\scriptscriptstyle L}$ in this case are played by the
quantities
\begin{equation}\label{42}
\;\;\;\;\;\beta_{term}=[\Ct \Fr]^{-1};\;\;\;\;\;\gamma_{term}=[\Ct
\Fr]
\end{equation}

In this case, the limits of the parameter~ $\beta_{term}$
as~$T\rightarrow \infty$ and~$T=0$ in the space of states correspond
to the limiting values~$\beta_L=0$ and~$\beta_L\rightarrow1$ in the
space of events. Thus, the state of the cold vacuum in the group of
Bogolyubov transformations corresponds to the \emph{proper inertial
reference system} in the special theory of relativity for the
group~$O(1,1)$ in the space of states. Other inertial reference
systems that are equivalent to the proper inertial reference system
correspond to different states of the arbitrary vacuum.

Thus, we are convinced once more that, among so far existing
theoretical approaches, only $(\hbar,k)$~dynamics [1] with
temperature-dependent complex wave functions provides the
fulfillment of the zeroth law at nonzero temperatures. Of course,
the possibility of constructing an adequate theory using the complex
density matrix remains open. At the same time, we note that the
authors of approaches to the description of thermal phenomena based
on real wave functions or real elements of the density matrix
understood intuitively the importance of the thermal equilibrium
condition in form~(24) as~$T\ne0.$ However, in the corresponding
theories, this condition always has the same trivial form~(31), (32)
or~(34)

\begin{equation} \label{44}
\;\;\;\;\; (\frac\hbar4)^2\left[\coth^2(\frac
12\beta\hbar\omega)-\left(\sinh^2(\frac{1}{2}\beta\hbar\omega)\right)^{-1}\right]\E_0
=\frac{\hbar\omega}{2}=(\frac\hbar4)^2,
\end{equation}
where the parameter~$\beta $ in both hyperbolic functions has the
same value~$ \frac{1}{kT}$. We see the principal difference between
formula~(44) and the following one
\begin{equation}\label{45}
\;\;\;\;\; (\frac\hbar4)^2\coth^2(\frac
{\hbar\omega}{2k_BT}=(\frac\hbar4)^2+(\frac\hbar4)^2\left(\sinh^2(\frac{\hbar\omega}{2k_BT_0})
\right)^{-1},
\end{equation}
in spite of their external similarity.

To obtain the desired result that allows treating the parameter~$T$
in formula~ (44) as a temperature, the authors of these papers
simply rewrote equality~(44=95) identically, carrying one of the
terms from the left to the right.   From the mathematical point of
view, this action is, of course, admissible. However, it is improper
from the physical point of view. It is very important to note that
the left- and right-hand sides of equality~(45=96) are related to
different physical entities: the characteristics of the system with
the temperature~$T$, which is able to fluctuate, is in the left-hand
side, and the characteristic of the stochastic environment with the
fixed temperature~$T_0$, which does not fluctuate because of an
infinitely large number of degrees of freedom of the quantum
thermostat, is in the right-hand side. They must be calculated
independently using the averaging over thermal CCSs.

That the above-mentioned formal procedure is inadmissible becomes
especially obvious in the limit of high temperatures when
formula~(44=95) transforms into the identity

\begin{equation}\label{46}
\;\;\;\;\; (\frac\hbar4)^2\coth^2(\frac
{\hbar\omega}{2k_{\scriptscriptstyle{B}}T})\rightarrow
(\frac\hbar4)^2\left(\sinh^2(\frac{\hbar\omega}{{2k_{\scriptscriptstyle{B}}T}})\right)^{-1}.
\end{equation}

\section*{\small Summary}
1. We have demonstrated that the physically significant subgroups of
the group of Bogolyubov $(u,v)$~transformations generate two types
of states: \emph{real squeezed} coherent states providing the
saturation of the Heisenberg "coordinate--momentum" UR at the zero
temperature and \emph{complex correlated coherent} states providing
the saturation of the Schr\"odinger "coordinate--momentum" UR at
finite temperatures. 2. On the microlevel, we have formulated a
quantum analogue of the zeroth law of stochastic thermodynamics in
the form of the saturated Schr\"odinger UR. 3. We have substantiated
the invariance of the saturated Schr\"odinger UR under the
Bogolyubov $u,v$~transformations. 4. We have proposed to regard the
saturated Schr\"odinger UR as an initial concept of quantum theory
at finite temperatures. 5. We have shown that in theories using real
wave functions or real elements of the density matrix, there is no
correlation between the coordinate and momentum fluctuations.
Therefore, in this case, the concept of thermal equilibrium
occurring as a result of the correlation between coordinate and
momentum fluctuations can be introduced only conditionally for the
states at the zero temperature, i.e., if there is no thermal
environmental influence. 6. We have established that in the theory
involving complex wave functions of the vacuum (thermal CCSs),
thermodynamics can be incorporated into quantum theory at any
temperatures (unlike the Umezawa thermofield dynamics).

 \begin{center}\small Reference
\end{center}

[1] Sukhanov A.D. and Golubjeva O.N.   Toward a quantum
generalization of equilibrium statistical thermodynamics:  $\hbar-k$
dynamics. Theoretical and Mathematical Physics, 160(2): 1177-1189
(2009)

[2] Sukhanov A.D.    A quantum generalization of equilibrium
statistical thermodynamics: effective macroparameters. Theoretical
and Mathematical Physics, 154(1): 153-164 (2008)

[3] Sukhanov A.D.and Golubjeva O.N. Arbitrary Vacuums as a Model of
Stochastic Environment: On the Problem of Incorporating
Thermodynamics into Quantum Theory. Physics of elementary particles
and atomic nuclei, Letters., Vol. 9, No. 3, pp. 303-311.(2012)

[4] Sukhanov A.D.and Golubjeva O.N. Squeezed coherent states as
thermal-like ones. ArXiv:1211.3017v1 [quant-ph] 13 Nov (2012)

[5] Umezawa H. Advanced Field Theory. Micro-, macro-, and Thermal
Physics. N.-Y.:AIP, (1993)

[6] J.M.Park. Improvement of Uncertainty Relation for Mixed States.
ArXiv:math-phys/0409008 v1, 3 Sept.(2004)

[7]  E.Wigner, M.M. Yanase. Inform. contents distributions.
Proc.Nat. Acad USA,  June, 49(6), 910 (1963)
\end{document}